\documentclass{article}
\usepackage{ijcai24}

\usepackage{times}
\usepackage{soul}
\usepackage{url}
\usepackage[hidelinks]{hyperref}
\usepackage[utf8]{inputenc}
\usepackage[small]{caption}
\usepackage{graphicx}
\usepackage{amsmath}
\usepackage{amsthm}
\usepackage{booktabs}
\usepackage{algorithm}
\usepackage{algorithmic}
\usepackage[switch]{lineno}
\usepackage{amssymb}

\title{Ultrasound SAM Adapter:
Adapting SAM for Breast Lesion Segmentation in Ultrasound Images}

\author{
Zhengzheng Tu$^1$\and
Le Gu$^2$\and
Xixi Wang$^{3}$\And
Bo Jiang$^4$\\
\affiliations
$^{1,2,3,4}$Anhui Provincial Key Laboratory of Multimodal Cognitive Computation, School of Computer Science and Technology, Anhui University, Hefei 230601, China\\
\emails
zhengzhengahu@163.com,
awngule@foxmail.com,
sissiw0409@foxmail.com,
zeyiabc@163.com
}

\begin{document}

\maketitle

\begin{abstract}
    Segment Anything Model (SAM) has recently achieved amazing results in the field of natural image segmentation. 
    However, it is not effective for medical image segmentation, owing to the large domain gap between natural and medical images. 
    In this paper, we mainly focus on ultrasound image segmentation. 
    As we know that it is very difficult to train a foundation model for ultrasound image data due to the lack of large-scale annotated ultrasound image data. 
    To address these issues, in this paper, we develop a novel Breast Ultrasound SAM Adapter, termed Breast Ultrasound Segment Anything Model (BUSSAM), which migrates the SAM to the field of breast ultrasound image segmentation by using the adapter technique. 
    To be specific, we first design a novel CNN image encoder, which is fully trained on the BUS dataset. 
    Our CNN image encoder is more lightweight, and focuses more on features of local receptive field, which provides the complementary information to the ViT branch in SAM. 
    Then, we design a novel Cross-Branch Adapter to allow the CNN image encoder to fully interact with the ViT image encoder in SAM module. 
    Finally, we add both of the Position Adapter and the Feature Adapter to the ViT branch to fine-tune the original SAM. 
    The experimental results on AMUBUS and BUSI datasets demonstrate that our proposed model outperforms other medical image segmentation models significantly. 
    Our code will be available at: \href{https://github.com/bscs12/BUSSAM}{https://github.com/bscs12/BUSSAM}.
\end{abstract}

\section{Introduction}
Segment Anything Model (SAM) \cite{kirillov2023segment} has received a lot of attention since it provides a powerful and versatile foundation model for natural image segmentation. 
Unlike traditional segmentation models, SAM does not need to be re-trained for a specific dataset, but can output corresponding segmentation results based on different prompts, such as points, boxes, text etc. 

Although SAM has achieved amazing results in the field of natural image segmentation. 
However, many studies have demonstrated that it usually performs poorly in the field of medical image (e.g., ultrasound image) segmentation tasks \cite{he2023accuracy,huang2023segment,deng2023segment,roy2023sam,mazurowski2023segment}. 
One main reason is the large domain gap between medical images and natural images. Thus, utilizing SAM to segment medical images directly does not fully take the advantage of the potential benefits of SAM's pre-trained on large-scale natural images. 
On the other hand, training large models for medical segmentation is usually difficult due to the lack of large-scale medical image segmentation datasets. 
To adapt SAM model for medical image segmentation problem, some works aim to utilize the fine-tuning strategy. 
For example, Ma and Wang et al. propose MedSAM \cite{ma2023segment} for the general medical image segmentation. 
MedSAM is trained on the carefully assembled datasets and can obtain desired performance. 
However, the assembled dataset is not large enough and also there is a modal imbalance in the training set, which limits the performance of MedSAM on ultrasound images. 
Zhang et al. propose SAMed \cite{zhang2023customized}, which applies a low-rank-based (LoRA) fine-tuning strategy in the SAM's image encoder.
Its encoder is fine-tuned on an annotated medical image segmentation dataset together with the prompt encoder and the mask decoder. 
Wu et al. propose MSA \cite{wu2023medical}, which enhances the ability of SAM to segment images by freezing the pre-trained parameters of SAM and inserting adapter modules at specific locations. 
For ultrasound images, 
Lin et al. propose SAMUS \cite{lin2023samus} 
which adapts SAM for ultrasound image segmentation task. 
It incorporates a CNN encoder branch together with cross-branch attention to fine-tune SAM. 
However, the cross-branch attention is high complexity which increases the burden of fine-tuning. 
Also, SAMUS uses a regular CNN backbone which fails to fully capture the subtle lesion cues for ultrasound images. 


In this paper, we mainly focus on breast ultrasound lesion segmentation task and develop a new Breast Ultrasound Segment Anything Model (BUSSAM) to migrate SAM model into the field of breast ultrasound image segmentation. 
Specifically, we first introduce an effective group attention guided CNN to fully capture the subtle visual salient cues for ultrasound image representation.
It aims to supplement the ViT encoder of SAM by exploiting some more information of local receptive fields.  
Then, we develop a new Cross-Branch Adapter which achieves the interaction between SAM and attentive CNN branch and also provides an adapter way to  fine-tune the SAM on ultrasound image data. 
Finally, 
we add a Position Adapter and a Feature Adapter to further fine-tune the ViT image encoder for SAM. 
The proposed BUSSAM can obtain more accurate segmentation results  while significantly reduces the deployment cost when compared to other related  methods.

Overall, the main contributions of this paper are summarized as follows, 

\begin{itemize}
\item We introduce a novel group attention guided CNN encoder to learn diverse features and capture the subtle salient cues for ultrasound images by focusing on local receptive features of ultrasound images. 
Moreover, it is lightweight and can be implemented very efficient.  

\item We design a novel simple Cross-Branch Adapter which allows the interaction between local CNN branch and ViT branch of SAM and also achieves a kind of fine-tuning for SAM model on ultrasound image dataset. 

\item We propose to introduce the Position Adapter and the Feature Adapter in SAM's ViT encoder branch to further fine-tune the parameters of ViT encoder. 

\item We evaluate the performance of the proposed method on two datasets. Experimental results  demonstrate that our proposed model significantly outperforms other related approaches. 
\end{itemize}

\section{Related Works}

\subsection{Parameter-Efficient Fine-Tuning}
Recently, foundation models have been growing rapidly in the fields of natural language processing (NLP) and computer vision (CV) \cite{devlin2018bert,brown2020language,radford2021learning,kirillov2023segment,wang2023seggpt}. These foundation models usually have billions, tens of billions or even larger number of parameters. In addition, a variety of parameter-efficient fine-tuning techniques \cite{li2021prefix,liu2021p,liu2023gpt} have arisen. Adapter tuning \cite{houlsby2019parameter} embeds an adapter module into the Transformer architecture, by fixing the parameters of the original pre-trained model and only fine-tuning the newly added adapter module during training. Adapter tuning can achieve the same effect as full fine-tuning with additional parameters with small size. 
Prompt tuning \cite{lester2021power} utilizes a singular prompt representation that is added to the embedded input. Beside of fewer parameters, prompt tuning enables the Transformer to modify the task representations in intermediate layers, contextualized by an input example. 
As the number of parameters in the pre-trained model increases, results of prompt tuning's approaches are close to results of full fine-tuning. In neural network models, model parameters are typically represented as matrices. 
LoRA \cite{hu2021lora,zhang2023adaptive,dettmers2023qlora} reduces the quantity of trainable parameters for downstream tasks by incorporating trainable rank decomposition matrices into every layer of the Transformer architecture.

\subsection{Medical Image Segmentation Using SAM}
Since the advent of SAM \cite{kirillov2023segment}, many works have attempted to apply SAM to medical image segmentation, and some have made encouraging progress. Ma and Wang et al. proposed MedSAM \cite{ma2023segment}, that is the first foundation model for generalized medical image segmentation. MedSAM is trained on large, well-assembled datasets, and its performance rivals or even surpasses that of professional models. Zhang and Liu et al. proposed SAMed \cite{zhang2023customized}, which applies a low-rank-based (LoRA) fine-tuning strategy to SAM's image encoder. The image encoder is fine-tuned on an annotated medical image segmentation dataset together with the prompt encoder and the mask decoder. Wu et al. proposed MSA \cite{wu2023medical}, which enhances the performance of SAM in segmenting images by freezing the pre-trained parameters of SAM and inserting adapters at specific locations. Lin et al. proposed SAMUS \cite{lin2023samus} which adapts SAM for ultrasound image segmentation task. It incorporates a CNN encoder branch together with cross-branch attention to fine-tune SAM. AutoSAM \cite{shaharabany2023autosam} replaces the prompt encoder of the original SAM with a customized prompt-generating network. This modification leads to state-of-the-art results in multiple medical image and video benchmarking tests without further fine-tuning of the SAM. Li et al. combine the advantages of both the foundation model and the domain-specific model to propose nnSAM \cite{li2023nnsam}, which combines the SAM model with the nnUNet model synergistically integrated to achieve more accurate and robust medical image segmentation.

\section{Methodology}

\begin{figure*}[t]
\centering
\includegraphics[width=1\linewidth]{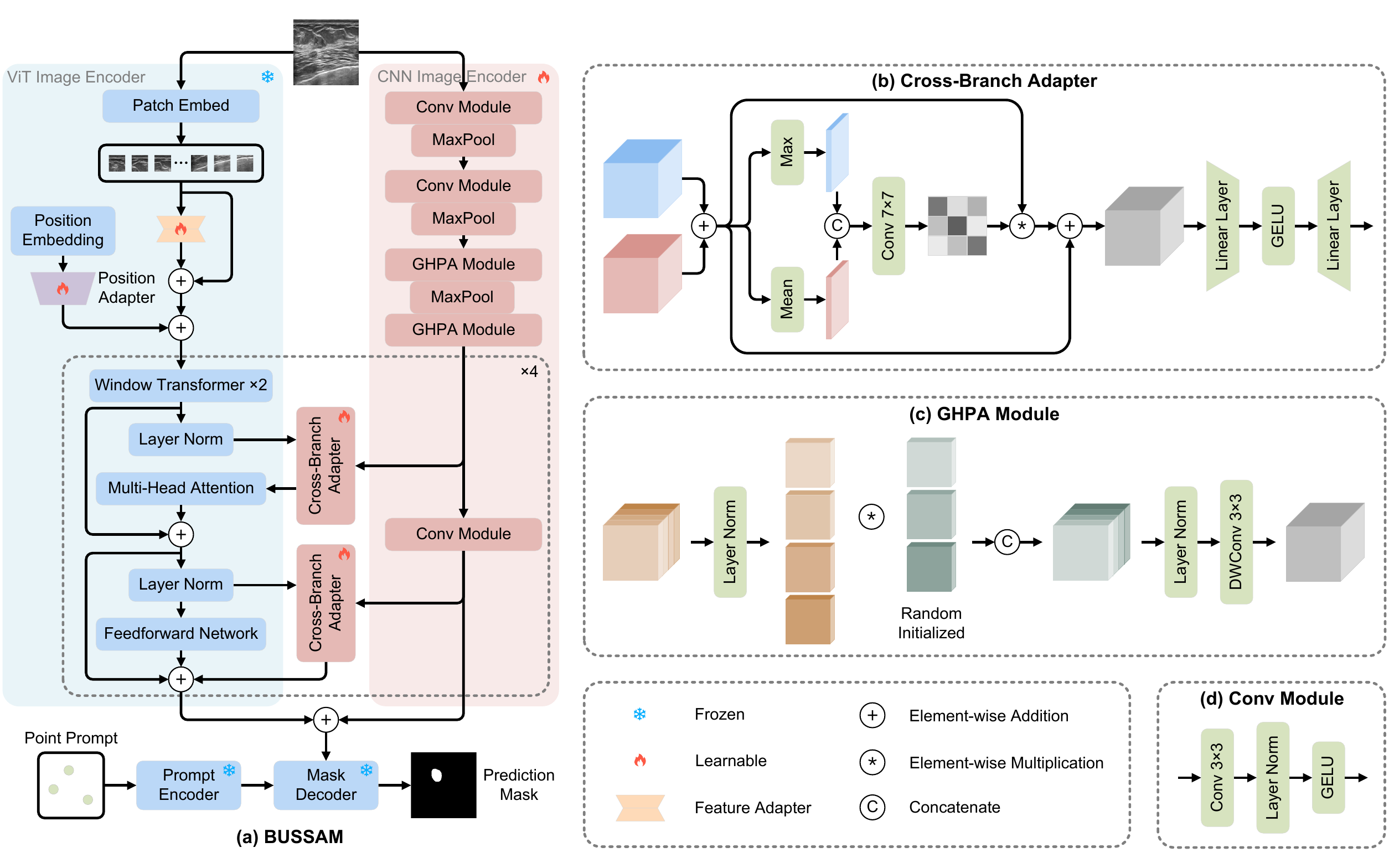}
\caption{Overview of our proposed Breast Ultrasound Segment Anything Model (BUSSAM) framework. 
}
\label{fig:overall_architecture}
\end{figure*}

In this section, we provide a detailed description of the proposed Breast Ultrasound Segment Anything Model (BUSSAM). 
The proposed BUSSAM framework is illustrated in Figure~\ref{fig:overall_architecture}, which mainly consists of CNN image encoder, Position Adapter in ViT image encoder and Cross-Branch Adapter.
Concretely, we first freeze all the parameters of Segment Anything Model (SAM).
Subsequently, we incorporate a CNN image encoder alongside the original SAM to capture information with localized receptive fields.
This CNN image encoder operates in parallel with ViT image encoder, both encoding the input image simultaneously.
Next, we introduce a novel Position Adapter module after the position embedding of ViT image encoder for fine-tuning.
Finally, we design a Cross-Branch Adapter to facilitate the interaction between CNN image encoder and ViT image encoder.
In the following sections, we delve into the specifics of CNN image encoder in Section~\ref{sec:CNN_Image_Encoder}, Position Adapter in ViT image encoder in Section~\ref{sec:position_adapter}, Cross-Branch Adapter in Section~\ref{sec:cross_branch_adapter}, and supervision strategy we adopt in Section~\ref{sec:supervision_strategy}.

\subsection{CNN Image Encoder}
\label{sec:CNN_Image_Encoder}
Considering the high complexity and diversity of boundaries and morphological features of breast lesion in ultrasound images, coupled with the presence of subtle structures such as lumps and cysts.
The segmentation process becomes notably challenging.
Therefore, to attain more precise lesion segmentation in breast ultrasound images, acquiring richer local information is particularly important.
To do this, we design a novel CNN image encoder that can complement the original ViT image encoder of SAM, collaborating to conduct deep encoding on the input image.
As shown in Figure~\ref{fig:overall_architecture}(a), the proposed CNN image encoder mainly consists of Convolution (Conv) module, Group multi-axis Hadamard Product Attention (GHPA) module and Global Max Pooling (GMP).

To be specific, an ultrasound image $X$ is inputted, we first alternate through two Conv modules and GAP operations and thus obtain the hidden features $F_{c} \in \mathbb{R}^{H_c \times W_c \times C_c}$, which can be defined as,
\begin{equation}
F_{c} = GAP(f_{\theta_2}(GAP(f_{\theta_1}(X))))
\label{equ:eq1}
\end{equation}
where $H_c$, $W_c$ and $C_c$ is the height, width and channel of hidden features.
$f_{\theta_1}(\cdot)$ and $f_{\theta_2}(\cdot)$ denote Conv module without shared parameters and Conv module is composed of a convolutional layer, a Layer Normalization (LN) and a GELU activation function, as is shown in Figure~\ref{fig:overall_architecture}(d). 
Then, we feed $F_{c}$ into two GHPA modules with a GAP operation to produce the enhanced image features $F^1_c \in \mathbb{R}^{H \times W \times C}$.
As depicted in Figure~\ref{fig:overall_architecture}(c), GHPA module consists of Layer Normalization (LN), Depthwise Separable Convolution (DWConv) and bilinear interpolation operation by following~\cite{ruan2023ege}.
Meanwhile, we perform a Conv module on $F^1_c$ to obtain the stronger image features $F^2_c \in \mathbb{R}^{H \times W \times C}$.
This process can be expressed as follows,
\begin{align}
\label{equ:eq2} & F^1_c = g_{\phi_2}(GAP(g_{\phi_1}(F_{c}))  \\
\label{equ:eq3} & F^2_c = f_{\theta_3}(F^1_c)
\end{align}
where $g_{\phi_1}(\cdot)$ and $g_{\phi_2}(\cdot)$ refer to two GHPA modules with different learnable parameters.
$H$, $W$ and $C$ denotes the height, width and channel number.
In this paper, unless otherwise stated, we set $H=W=32$ and $C=768$ in all experiments.

Through experimental demonstrations, we illustrate that the intergration of Conv module and GHPA module not only substantially reduces the computational resource, but also outperforms the standalone convolutional layer in terms of performance. 
This enhancement is attributed to the capability of GHPA module to effectively weight features from diverse perspectives, resulting in a better feature representation compared to conventional convolution. 
To enable multi-scale feature extraction, we employ maximum pooling to diminish the feature resolution.

\subsection{Position Adapter in ViT Image Encoder}
\label{sec:position_adapter}
In order to promote the development of image segmentation field, Kirillov et al.~\cite{kirillov2023segment} propose a
Segment Anything Model (SAM), which is a foundation model trained on a large SA-1B dataset with 1 billion masks and 11 million natural images.
SAM mainly contains ViT image encoder, prompt encoder and mask decoder.
ViT image encoder processes the input image and extracts its features.
These features are then combined with the prompts from the prompt encoder.
The mask decoder takes this combined information and generates the segmented masks for the input image. 
To enhance the applicability of SAM in the realm of ultrasound image, we have improved its core components.
As shown in Figure~\ref{fig:overall_architecture}(a), we first add a Feature Adapter as suggested in~\cite{lin2023samus} to skip and augment the feature representation capabilities of ultrasound images.
Furthermore, considering that the location of breast lesions in ultrasound images is crucial for segmentation, in this paper, we introduce a novel Position Adapter based on the original ViT image encoder.
This adapter facilitate the lesion understanding and localization by encoding and extracting more precise position.


For Position Adapter, let $P$ denotes the position embedding. 
Specifically, we first downsample $P$ using Maximum Pooling on the channel dimension, and then employ Group multi-axis Hadamard Product Attention (GHPA) module to process it. 
This process can be formulated as,
\begin{equation}
\tilde{P} = \sigma_1(GN(g_{\phi_3}(MP(P))))
\label{eq:eq4}
\end{equation}
where $MP(\cdot)$ denotes Maximum Pooling with kernel size 2 and stride 2.
$g_{\phi_3}(\cdot)$ denotes Group multi-axis Hadamard Product Attention module with learnable parameter.
$GN(\cdot)$ and $\sigma_1(\cdot)$ represent Group Normalization and GELU activation function, respectively.

By incorporating the proposed Position Adapter, we just update the parameters of its and feature adapter for SAM, while keeping everything else frozen.
This method enables us to seamlessly tailor SAM to medical image tasks with minimal adjustments. 


\subsection{Cross-Branch Adapter}
\label{sec:cross_branch_adapter}
To establish information interaction between CNN image encoder and ViT image encoder, in this paper, we design a Cross-Branch Adapter module.
According to the above, given $F_c \in \mathbb{R}^{H \times W \times C}$ and $F_v \in \mathbb{R}^{H \times W \times C}$ as the feature representations of CNN image encoder and ViT image encoder, respectively.
As shown in Figure~\ref{fig:overall_architecture}(b), we first sum $F_c$ and $F_v$ and then we obtain $F_{max}$ and $F_{mean}$ by performing Maximum Pooling (MP) and Average Pooling (AP) on the channel dimensions respectively, which can be denoted as,
\begin{align}
\label{equ:eq5} & H_{max} = MP(F_v + F_c)  \\
\label{equ:eq6} & H_{mean} = AP(F_v + F_c)
\end{align}
where $MP(\cdot)/AP(\cdot)$ indicates the maximum/average pooling operation on the channel dimension.
Subsequently, We concatenate $H_{max}$ and $H_{mean}$ and then process them using a convolutional layer with kernel size $7 \times 7$, followed by a sigmoid activation function to derive the spatial attention weight $W$. 
This weight $W$ is then used to modulate the spatial attention of the fused features of $F_v$ and $F_c$, and the output is obtained through two linear layers and a GELU activation function. 
To regulate the weight of Cross-Branch Adapter, we finally add a scaling factor to adjust the output. 
The whole process can be formulated as follows,
\begin{align}
\label{equ:eq7} & W = \sigma_2(f(H_{max} \parallel H_{mean}))  \\
\label{equ:eq8} & \widetilde{H} = W \odot (F_v + F_c)  \\
\label{equ:eq9} & H = \alpha h_2(\sigma_3(h_1(\widetilde{H})))
\end{align}
where $\parallel$ denotes the concatenation operation along the channel dimension.
$f(\cdot)$ refers to a convolutional layer with kernel size $7 \times 7$.
$\sigma_2(\cdot)$ represents sigmoid activation function.
$\odot$ denotes element-wise multiplication.
$h_1(\cdot)$ signifies a linear layer with a decreasing number of features.
$\sigma_3(\cdot)$ denotes GELU activation function.
$h_2(\cdot)$ indicates a linear layer with an increasing number of features.
$\alpha$ denotes scaling factor.

\begin{table*}[t]
\caption{Quantitative evaluation results on AMUBUS dataset. $\uparrow$ means higher is better and $\downarrow$ means lower is better. The optimal outcomes are denoted using \textbf{bold} typeface.}
\label{table3}
\centering
\begingroup
\setlength{\tabcolsep}{10pt}
\renewcommand{\arraystretch}{1.5}
\begin{tabular}{c|ccccc}
\toprule
\textbf{Method} & \textbf{Acc (\%) ↑} & \textbf{Se (\%) ↑} & \textbf{Dice (\%) ↑} & \textbf{IoU (\%) ↑} & \textbf{HD (mm) ↓} \\
\midrule
U-Net \cite{ronneberger2015u}          & 97.67 & 86.18 & 80.38 & 70.80 & 6.68 \\
SegNet \cite{badrinarayanan2017segnet} & 98.93 & 87.34 & 82.90 & 72.71 & 6.73 \\
DeepLabV3+ \cite{chen2018encoder}      & 96.52 & 84.51 & 79.25 & 68.90 & 7.17 \\
U-Net++ \cite{zhou2018unet++}          & 98.19 & 87.60 & 83.56 & 73.36 & 6.43 \\
PraNet \cite{fan2020pranet}            & 98.90 & 86.82 & 83.00 & 72.83 & 6.85 \\
RF-Net \cite{wang2021residual}         & 98.14 & 86.63 & 83.27 & 73.09 & 6.68 \\
TransResUnet \cite{tomar2022transresu} & 98.95 & 87.65 & 83.74 & 74.62 & 6.38 \\
SAMUS \cite{lin2023samus}              & 99.29 & 88.52 & 85.89 & 76.36 & 6.16 \\
\textbf{BUSSAM (Ours)} & \textbf{99.32} & \textbf{89.16} & \textbf{86.59} & \textbf{77.21} & \textbf{6.14} \\
\bottomrule
\end{tabular}
\endgroup
\end{table*}

\begin{table*}[t]
\caption{Quantitative evaluation results on BUSI dataset. $\uparrow$ means higher is better and $\downarrow$ means lower is better. The optimal outcomes are denoted using \textbf{bold} typeface.}
\label{table4}
\centering
\begingroup
\setlength{\tabcolsep}{10pt}
\renewcommand{\arraystretch}{1.5}
\begin{tabular}{c|ccccc}
\toprule
\textbf{Method}   & \textbf{Acc (\%) ↑} & \textbf{Se (\%) ↑} & \textbf{Dice (\%) ↑} & \textbf{IoU (\%) ↑} & \textbf{HD (mm) ↓} \\
\midrule
U-Net \cite{ronneberger2015u}          & 97.12 & 88.56 & 84.61 & 75.16 & 8.44 \\
SegNet \cite{badrinarayanan2017segnet} & 96.53 & 87.56 & 82.37 & 73.58 & 8.62 \\
DeepLabV3+ \cite{chen2018encoder}      & 95.70 & 86.46 & 81.67 & 71.29 & 9.63 \\
U-Net++ \cite{zhou2018unet++}          & 97.12 & 89.09 & 85.70 & 76.11 & 8.46 \\
PraNet \cite{fan2020pranet}            & 96.83 & 88.11 & 85.68 & 76.36 & 8.71 \\
RF-Net \cite{wang2021residual}         & 96.74 & 88.62 & 86.52 & 78.24 & 8.41 \\
TransResUnet \cite{tomar2022transresu} & 97.13 & 89.72 & 86.13 & 77.61 & 8.50 \\
SAMUS \cite{lin2023samus}              & 97.56 & 89.82 & 88.89 & 79.36 & 8.35 \\
\textbf{BUSSAM (Ours)} & \textbf{98.06} & \textbf{91.49} & \textbf{89.95} & \textbf{82.31} & \textbf{8.27} \\
\bottomrule
\end{tabular}
\endgroup
\end{table*}

\begin{table*}[t]
\caption{Results of the ablation study on AMUBUS dataset for the proposed modules in BUSSAM. CNN denotes CNN image encoder, PosA denotes Position Adapter and CBA denotes Cross-Branch Adapter.
$\uparrow$ means higher is better and $\downarrow$ means lower is better.}
\label{table5}
\centering
\begingroup
\setlength{\tabcolsep}{10pt}
\renewcommand{\arraystretch}{1.5}
\begin{tabular}{ccc|ccccc}
\toprule
\textbf{CNN} & \textbf{PosA} & \textbf{CBA} & \textbf{Acc (\%) ↑} & \textbf{Se (\%) ↑} & \textbf{Dice (\%) ↑} & \textbf{IoU (\%) ↑} & \textbf{HD (mm) ↓} \\
\midrule
           &            &            & 98.69 & 87.49 & 78.88 & 67.38 & 6.39 \\
\checkmark &            &            & 99.10 & 87.64 & 84.50 & 74.12 & 6.35 \\
           & \checkmark &            & 99.17 & 87.82 & 84.37 & 74.18 & 6.32 \\
\checkmark & \checkmark &            & 99.30 & 88.61 & 86.21 & 76.47 & 6.15 \\
\checkmark & \checkmark & \checkmark & 99.32 & 89.16 & 86.59 & 77.21 & 6.14 \\
\bottomrule
\end{tabular}
\endgroup
\end{table*}

\begin{table*}[t]
\caption{Results of the ablation study on BUSI dataset for the proposed modules in BUSSAM. CNN denotes CNN image encoder, PosA denotes Position Adapter and CBA denotes Cross-Branch Adapter. $\uparrow$ means higher is better and $\downarrow$ means lower is better.}
\label{table6}
\centering
\begingroup
\setlength{\tabcolsep}{10pt}
\renewcommand{\arraystretch}{1.5}
\begin{tabular}{ccc|ccccc}
\toprule
\textbf{CNN} & \textbf{PosA} & \textbf{CBA} & \textbf{Acc (\%) ↑} & \textbf{Se (\%) ↑} & \textbf{Dice (\%) ↑} & \textbf{IoU (\%) ↑} & \textbf{HD (mm) ↓} \\
\midrule
           &            &            & 96.99 & 86.79 & 83.88 & 73.90 & 8.61 \\
\checkmark &            &            & 97.33 & 88.97 & 88.25 & 78.76 & 8.39 \\
           & \checkmark &            & 97.56 & 89.82 & 88.89 & 79.36 & 8.35 \\
\checkmark & \checkmark &            & 97.88 & 90.52 & 89.23 & 81.67 & 8.30 \\
\checkmark & \checkmark & \checkmark & 98.06 & 91.49 & 89.95 & 82.31 & 8.27 \\
\bottomrule
\end{tabular}
\endgroup
\end{table*}

\subsection{Supervision Strategy}
\label{sec:supervision_strategy}
In this paper, the whole network is trained in an end-to-end manner.
During training, we use Binary Cross Entropy (BCE) loss $\mathcal{L}_{BCE}$ and Dice loss $\mathcal{L}_{Dice}$ to supervise the whole network.
They can be defined as,
\begin{equation}
\begin{split}
\mathcal L_{BCE} = &- \sum_{x=1}^{H} \sum_{y=1}^{W} \left[ G(x,y) \log(P(x,y)) \right. \\
&\left. + (1-G(x,y)) \log(1-P(x,y)) \right] 
\end{split}
\label{eq:eq10}
\end{equation}
\begin{equation}
\mathcal{L}_{Dice} = 1 - \frac{2TP+1}{2TP+FN+FP+1}
\label{eq:eq11}
\end{equation}
where $G(x,y)$ represents the ground-truth of input ultrasound image, while $P(x, y)$ represents the prediction map.
TP, FN, and FP represent the true positive, false negative, and false positive, respectively, in relation to the comparison between the ground truth and the prediction map.

Finally, the total loss can be formulated as follow,
\begin{equation}
\mathcal{L} = \beta \mathcal{L}_{BCE} + (1-\beta) \mathcal{L}_{Dice},
\end{equation}
where $\beta \in [0, 1]$ is a balanced hyper-parameter.
In our experiment, we empirically set $\beta=0.2$.


\section{Experiments}
In this section, we validate and demonstrate the effectiveness of the proposed BUSSAM through extensive experiments.
Specifically, we begin by presenting the datasets and evaluation metrics we utilize in Section~\ref{Datasets}.
Next, we describe our experimental setup in Section~\ref{Experiments Settings}.
Then, 
we compare our method experimentally with other methods from both quantitative and qualitative perspectives in Section~\ref{Quantitative Evaluation}.
Finally, we conduct ablation experiments on our designed modules in Section~\ref{Ablation Study}.


\subsection{Datasets and Evaluation Metrics}\label{Datasets}
\textbf{Datasets.}
In this work, we use AMUBUS dataset and BUSI~\cite{al2020dataset} dataset for conducting the experiments. 
AMUBUS dataset comprises 2642 ultrasound images, with 2113 images in the training set and 529 images in the testing set. 
These images are acquired from 528 patients at the First Affiliated Hospital of Anhui Medical University using the Resona 7 and Toshiba 660a ultrasound systems. 
No cropping or data cleaning is applied to these images, and they include a significant number of challenge samples.
The dataset contains images of various sizes, ranging from $256 \times 256$ to $1072 \times 756$.
For ease of result organization, all the images are scaled to $256 \times 256$.
To comprehensively validate the performance of BUSSAM, we also conduct experiments on BUSI dataset. The samples in BUSI dataset are collected from 600 female patients aged between 25 and 75 years. 
This dataset contains 780 images (including 133 normal samples) collected from Baheya Hospital using LOGIQ E9 and LOGIQ E9 Agile ultrasound systems. 
The average size of the images in BUSI is $500 \times 500$. 
We specifically select samples with tumor targets present in BUSI and exclude samples containing multiple tumor targets. 
The training set and test set are then divided according to a ratio of 4 to 1.
The ratio of benign or malignant samples in the training set to the test set is also 4 to 1.

\textbf{Evaluation Metrics.}
We employ five test metrics to comprehensively evaluate the performance of our method. These metrics include Accuracy, Sensitivity, Dice index~\cite{taha2015metrics}, Intersection-Over-Union and Hausdorff Distance. Accuracy (Acc) represents the ratio of correctly categorized pixels to the total number of pixels in the segmentation result. Sensitivity (Se), also known as true positive rate and recall, measures the ability to accurately segment the region of interest in a segmentation experiment. Dice index (Dice), also known as the overlap index, quantifies the ratio of the area where two objects intersect to the total area. It is commonly used to measure the similarity or overlap of two samples. Intersection-Over-Union (IoU), which is also known as Jaccard's index and assess the similarity or overlap between two sets. Hausdorff Distance (HD) is a metric utilized to quantify the similarity between two sets of point sets. It is commonly employed in segmentation tasks to assess the accuracy of boundary segmentation.

\begin{figure*}[t]
\centering
\includegraphics[width=1\linewidth]{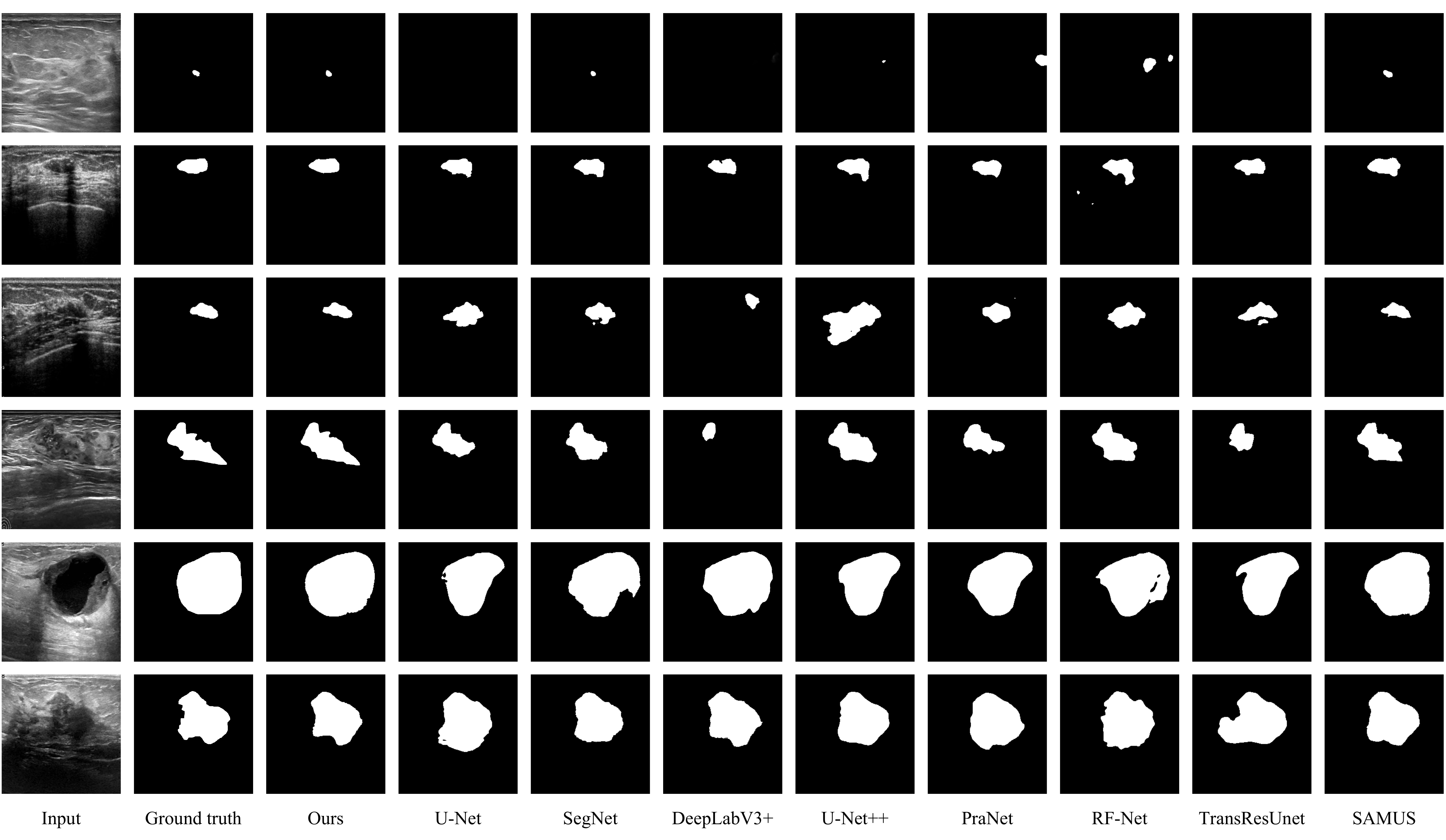}
\caption{Qualitative comparison results between the proposed BUSSAM method and other SOTA methods. Rows one through four correspond to samples from the AMUBUS dataset, while rows five through six represent samples from the BUSI dataset. It is recommended to view the images in color mode for optimal visualization.}
\label{fig:visualization}
\end{figure*}

\subsection{Experiments Settings}\label{Experiments Settings}
The proposed BUSSAM model is deployed on an NVIDIA GeForce RTX 3090 GPU, and all experiments are conducted using the publicly available PyTorch 1.8.0 platform and Python 3.8 environment.
We use SAM's ViT-B pre-training weight to initialize the proposed BUSSAM network and freeze all parameters during training. 
During the training process, we only update the parameters of CNN image encoder, Position Adapter, Feature Adapter and Cross-Branch Adapter, while the remaining parameters remain frozen.
%
%
To be concrete, each image is initially resized to $256 \times 256$, and various data augmentation methods such as normalization, random cropping, and random flipping are employing to augment the images. 
The batch size is set to 8, and the maximum number of epochs is set to 100.
Moreover, we use AdamW optimizer \cite{loshchilov2017decoupled} with weight decay 0.1 to train our network.
The initial learning rate is set to 5e-4.
We add warm up and linear decay strategies to adjust the learning rate.

\subsection{Quantitative Evaluation}\label{Quantitative Evaluation}
We compare the proposed BUSSAM with several leading segmentation methods, including U-Net \cite{ronneberger2015u}, SegNet \cite{badrinarayanan2017segnet}, DeepLabV3+ \cite{chen2018encoder}, U-Net++ \cite{zhou2018unet++}, PraNet \cite{fan2020pranet}, RF-Net \cite{wang2021residual}, TransResUnet \cite{tomar2022transresu}, and SAMUS \cite{lin2023samus}. 
U-Net \cite{ronneberger2015u}, SegNet \cite{badrinarayanan2017segnet} and DeepLabV3+ \cite{chen2018encoder} are among the most widely used segmentation baselines. 
U-Net++ \cite{zhou2018unet++} propose a redesigned skip-connection strategy to consider information from multi-level features.
PraNet \cite{fan2020pranet} use reverse attention to refine the boundaries of segmented tumors. 
RF-Net \cite{wang2021residual} introduce a residual feedback strategy to focus more attention on cluttered regions. 
TransResUnet \cite{tomar2022transresu} combine Transformer with CNNs to consider both local information and global information.
SAMUS \cite{lin2023samus} adapts SAM for ultrasound image segmentation task through fine-tuning.

To ensure a fair comparison, all the aforementioned models are retrained on both AMUBUS dataset and BUSI~\cite{al2020dataset} dataset. 
Table \ref{table3} presents the quantitative results of our proposed BUSSAM method in comparison to eight other methods on AMUBUS dataset. 
Our method achieves scores of 99.32\%, 89.16\%, 86.59\%, 77.21\%, and 6.14 mm on the five metrics, outperforming the second-place method by 0.03\% for Acc, 0.64\% for Se, 0.70\% for Dice and 0.85\% for IoU, respectively.
Additionally, it reduces the HD by 0.02 mm.
On the BUSI dataset, as presented in Table \ref{table4}, our proposed BUSSAM method achieves scores of 98.06\%, 91.49\%, 89.95\%, 82.31\%, and 8.27 mm for the five metrics,
improving Acc, Se, Dice, and IoU by 0.50\%, 1.67\%, 1.06\%, and 2.95\%, respectively, while reducing HD by 0.08 mm, compared to the second place finisher. 
These results clearly demonstrate the superior performance of the proposed BUSSAM model, indicating its capability to significantly enhance the segmentation performance.

In addition, in order to provide a more intuitive comparison, we visualize our proposed BUSSAM model alongside the other SOTA methods separately.
Figure \ref{fig:visualization} displays a visual comparison between the proposed BUSSAM method and the other SOTA methods. 
The visual comparison reveals that the segmentation results produced by our proposed BUSSAM closely resemble the ground-truth. 
Moreover, to further validate the effectiveness of our proposed BUSSAM, we conduct visual analysis for various challenging cases, such as small tumors in the first row, noise interference in the third row, and boundary blurring in the sixth row. 
It is evident that, when compared to other SOTA methods, our proposed BUSSAM consistently generates more accurate saliency maps in these challenging scenarios. 
In a word, these visualization results further underscore the capability of the proposed BUSSAM model to address breast lesion segmentation under various complex challenges.

\subsection{Ablation Study}\label{Ablation Study}
We evaluate the effectiveness of different modules in the proposed BUSSAM model through ablation experiments on the AMUBUS dataset and BUSI dataset. 
As depicted in Table \ref{table5} and \ref{table6}, the first row represents the baseline result, which corresponds to the original SAM.
It is evident that the advantages of SAM cannot be fully utilized.
Firstly, we add the Position Adapter module 
after the position embedding of the original SAM, resulting in an enhancement compared to the baseline method.
This demonstrates that the proposed Position Adapter module can improve SAM's generalization ability to BUS images. 
Subsequently, we introduce a novel CNN image encoder and allow ViT image encoder to interact with it through simple summation, leading to further improvement.
This indicates that our designed CNN image encoder can significantly enhance segmentation performance. 
Following this, we incorporate the Position Adapter and CNN image encoder simultaneously, yielding better results than their respective individual cases.
Finally, by introducing a novel Cross-Branch Adapter module instead of simple summation, we achieve a further boost, demonstrating that the proposed Cross-Branch Adapter module can enhance the information interaction between CNN image encoder and ViT image encoder.


\section{Conclusion}\label{Conclusion}
In this work, we develop a novel Breast Ultrasound SAM Adapter, termed Breast Ultrasound Segment Anything Model (BUSSAM), which migrates SAM to the field of breast ultrasound image segmentation tasks by using adapter technique. Specifically, we design a novel CNN image encoder that is learnable and we let it be fully trained on the BUS dataset. Compared with the ViT image encoder, the CNN image encoder is more lightweight and focus more on local receptive field features which provides the complementary information to the ViT branch in SAM. Second, we design a Cross-Branch Adapter in order to allow the CNN image encoder to fully interact with the ViT image encoder. Finally, we add a Position Adapter and a Feature Adapter to the ViT branch to fine-tune the original SAM. The experimental results on AMUBUS dataset and BUSI dataset show that our model is significantly better than other medical image segmentation models.

\bibliographystyle{named}
\bibliography{ijcai24}

\end{document}